\documentclass[%
 reprint,
superscriptaddress,
showpacs,
 amsmath,amssymb,
 aps,
]{revtex4-1}

\usepackage{graphicx}
\usepackage{color}
\usepackage{verbatim}
\usepackage{subfigure}
\usepackage{multirow}
\usepackage{braket}

\definecolor{Red}{rgb}{1,0,0}

\begin{document}


\title{\ \\ \ \\ \ \\
Quantum projectors implemented with  optical directional  couplers \\ fabricated by  Na/K ion-exchange  in soda-lime glass} 

\author{Xes\'us Prieto-Blanco} 
\email{Corresponding author: suso.linares.beiras@usc.es}
\affiliation{Quantum Materials and Photonics Research Group, Optics Area, Applied Physics Department, Faculty of Physics / Faculty of Optics and Optometry, Universidade de  Santiago de Compostela, Campus Vida s/n, E-15782, Santiago de Compostela, Galicia, Spain}
\author{Carlos Montero-Orille}
\affiliation{Quantum Materials and Photonics Research Group, Optics Area, Applied Physics Department, Faculty of Physics / Faculty of Optics and Optometry, Universidade de  Santiago de Compostela, Campus Vida s/n, E-15782, Santiago de Compostela, Galicia, Spain}
\author{Jes\'us Li\~nares}
\affiliation{Quantum Materials and Photonics Research Group, Optics Area, Applied Physics Department, Faculty of Physics / Faculty of Optics and Optometry, Universidade de  Santiago de Compostela, Campus Vida s/n, E-15782, Santiago de Compostela, Galicia, Spain}
\author{H\'ector Gonz\'alez-N\'u\~nez}
\affiliation{Quantum Materials and Photonics Research Group, Optics Area,  Applied Physics Department, Faculty of Physics / Faculty of Optics and Optometry, Universidade de Santiago de Compostela, Campus Vida s/n, E-15782, Santiago de Compostela, Galicia, Spain}
\author{Daniel Balado}
\affiliation{Quantum Materials and Photonics Research Group, Optics Area,  Applied Physics Department, Faculty of Physics / Faculty of Optics and Optometry, Universidade de  Santiago de Compostela, Campus Vida s/n, E-15782, Santiago de Compostela, Galicia, Spain}

\begin{abstract} {We present a preliminary theoretical and experimental study of quantum projectors implemented by integrated optical directional couplers fabricated by  ion-exchange Na/K processes in soda-lime glass.  Theoretical considerations about devices formed by concatenated 2x2 directional couplers are presented in order to show their capabilities for implementing $N$-dimensional quantum projective measurements,  and concomitantly the production of  1-qudit states. Since the  fundamental unit of these devices are 2x2 directional couplers, we present an experimental study for obtaining, by an optical characterization, empiric relationships between fabrication and optical parameters of such couplers.   Likewise, a two-dimensional quantum projector is demonstrated in such a way that projective measurements are obtained for the states of  $X$ (diagonal) and $Y$ (circular) bases.}

   \end{abstract}

\maketitle 

\section{Introduction}
Projective measurements are the most common way to detect  quantum states, in general, and quantum light states, in particular. It is an important task in most of quantum  optical circuits intended for quantum computation, quantum cryptography and in general quantum processing \cite{Fox,AgarwalBook}. In the case of quantum communications, quantum light states propagate in optical fibers or free space and can be  excited in codirectional  optical modes as those ones of Multi-Core optical Fibers (MCF) \cite{Overview,Canas2017, Balado2019},  or in collinear optical modes  as those ones of Few Mode optical Fibers (FMF) \cite{Bai12,BaladoJMO} and free space.   On the other hand, codirectional modes coming from MCFs can be easily coupled to an integrated  optical circuit (IOC), therefore, such circuits are ideal candidates to perform quantum transformations, in particular, quantum projective measurements. We call this type of circuits as  integrated quantum optical circuits (IQOCs). In addition, this solution can be extended to collinear modes, for that, a prior  spatial demultiplexing process, along with the use modal converters, is required in order to get  a spatial separation of optical modes what  facilitates their coupling  into an IQOC. Accordingly the implementation of the mentioned quantum transformations and/or projective measurements can be made.  Alternatively the use of photonic lanterns can implement such operations in a compact  form \cite{Leon2013}. As commented, these quantum circuits can perform different transformations but  projective measurements are usually required in the final part of quantum communications systems (Bob system). 

In this work we focus on these projective measurements used for quantum cryptography in optical fibers\cite{Fox, Canas2017, Balado2019}, although other applications are also possible, as quantum simulations, quantum sensors, and so on.  IQOCs can be  realized with integrated optical elements as single mode channel waveguides (SMW), directional couplers and integrated phase shifters \cite{Lee,Najafi92,SPIE2000, Righini1997}.  Accordingly, integrated quantum optical projectors (IQOP) can be implemented with these integrated optical elements, and  therefore an experimental realization of these projectors is an important  and relevant task.   Several platforms can be used, as for example,  silicon-on-insulator (as SiO$_{2}$), Lithium Niobate, Gallium Arsenide  and so on \cite{Politi2009,Wang2020}. However, to our knowledge, ion-exchanged glass (IExG) platform has not be considered as a integration technology for quantum photonics, although IExG has given important results in classical and semiclassical  photonics and has been suggested several years ago \cite{Linares2011}. We propose the experimental use of this platform for implementing IQOPs, although many other quantum components could be implemented, as for example, quantum simulators \cite{boson},  quantum random walk  and so on. These implementations will be efficient  as long as a few qubits are considered, which is related to the well-known scalability difficulty of quantum integrated photonics. In particular, we propose quantum projectors implemented with  concatenated optical directional  couplers and phase shifters fabricated by  Na/K ion-exchange processes in soda-lime glass. Since these projectors are intended  for quantum cryptography purposes, they are designed in a such  way that a quantum random choice of measurement basis is realized \cite{Balado2019}. The fundamental or basic unit of these projectors is the directional coupler, therefore  we present the results of fabrication  and the optical tests of such couplers. In particular, we make a semiclassical test of projective measurements of the quantum states belonging to   bases X and Y. The results show that IExG is a platform with enough advantages to be used in quantum technology along with other platforms. The plan of the work is the following. In section~\ref{sectionII} we present the theoretical fundamentals on passive quantum projectors with random choice of measurement basis,  as required in the most of quantum cryptography protocols. In particular, we make clear this kind of projectors with the case $N=2$,  although their generalization to arbitrary dimension is quite straightforward. Next, in section~\ref{sectionIII} the fabrication of a 2x2 projector by IExG is described, in particular by a  Na/K ion-exchange process in soda-lime glass, and by paying special attention to the experimental  curves obtained to  determine the fabrication parameters of the circuit. In section~\ref{sectionIV} optical test of a basic projector  2x2 is presented, and in particular, we use an external optical grating to change the basis of quantum states and show, in a semiclassical way, that projective measurements can be realized.  Finally in section~\ref{sectionV} conclusions are presented.

\section{Quantum projectors based on directional couplers}
\label{sectionII}
The  quantum light states with a major  interest  are the single photon states, which are, for instance, used in quantum cryptography, and   in particular 1-qubits and 1-ququart states. In fact,  1-ququart states are the states giving rise to the beginning of the so called high dimension quantum cryptography with 1-qudits ($N=d>2$),  that is,
\begin{equation}
\vert L\rangle=\sum_{j=1}^{d}c_{j}\vert 000 ...1_{j}...000\rangle, 
\end{equation}
where the states have to belong to  the so called mutually unbiased bases (MUBs), that is, $\vert c_{1}\vert =...=\vert c_{j}\vert =...=\vert c_{d}\vert $. A source of single photons can be obtained from SPDC sources \cite{AgarwalBook} or even from lasers, that is, from multimode  coherent states $\vert \alpha_{1}...\alpha_{N}\rangle$ with a very low photonic excitation (weak coherent states), that is, $\vert \alpha_{j}\vert\ll1$, therefore
\begin{equation}
\vert L\rangle\approx \vert 0\rangle+ \sum_{j=1}^{d}\alpha_{j}\vert 000 ...1_{j}...000\rangle. \end{equation}
Let us start by considering the case of a  two-dimensional quantum problem, that is, $N=2$, therefore, 1-qubits states will be used. This case allows us to make clear how to implement integrated quantum optical projectors. 
As commented, the fundamental units of these devices are the synchronous directional couplers along with integrated phase shifters. It is well known that synchronous directional couplers can be represented by the following matrix
\begin{equation}\label{Xcoupler}
X_{\theta}=\begin{pmatrix} \cos\theta& i\sin\theta \\
 i\sin\theta& \cos\theta \end{pmatrix},
\end{equation}
where $\theta=\kappa L$, with $\kappa$ the coupling coefficient and $z=L$  the coupling effective length  of the coupler.  For a two-dimensional IQOP the   required couplers are by one hand a  $X_{\pi/4}$, that is, 3dB (or 1:1) coupler,  which in turn corresponds to the logic gate \smash{$\sqrt{X}$}, and on the other hand a $X_{\pi/2}$ corresponding to the usual logic gate $X$. Moreover a phase shifters are required, that is 
\begin{equation}
Z_{\Phi}=\begin{pmatrix} e^{-i\Phi/2}& 0 \\
 0& e^{i\Phi/2}\end{pmatrix}.
\end{equation}
 Phase shifters $Z_{\pi/2}$ and $Z_{\pi}$ are usually needed. Therefore,  with these devices $X_{\theta}$ and $Z_{\Phi}$ concatenated  we can implement IQOPs. For example, let us consider the bases X and Y with the following single photon states excited in two channel waveguides $j$ and $j'$,
\begin{equation*}
{\rm X}= \{\vert 1_{D}\rangle, \vert 1_{A}\rangle\}=
\end{equation*}
\vspace{-1,35cm}
\ \\
\begin{equation}\label{baseX}
=\{\frac{1}{\sqrt{2}}(\vert 1_{j}\rangle+\vert 1_{j'}\rangle), \frac{1}{\sqrt{2}}(\vert 1_{j}\rangle-\vert 1_{j'}\rangle)\},
\end{equation}
\begin{equation*}
{\rm Y}= \{\vert 1_{L}\rangle, \vert 1_{R}\rangle\}=\\
\end{equation*}
\vspace{-1,35cm}
\ \\
\begin{equation}\label{baseY}
=\{\frac{1}{\sqrt{2}}(\vert 1_{j}\rangle+i\vert 1_{j'}\rangle), \frac{1}{\sqrt{2}}(\vert 1_{j}\rangle-i\vert 1_{j'}\rangle)\}.
\end{equation}

If these channel waveguides are coupled by a coupler $P_{Y}$$=$$X_{\pi/4}$ then a IQOP is implemented for the basis Y, that is, the input state $\vert 1_{L}\rangle$ becomes the state  $i\vert 1_{j'}\rangle$ at the output, and the state $\vert 1_{D}\rangle$ becomes the state  $i\vert 1_{j}\rangle$ at the output. On the other hand, the IOQC defined by $P_{X}$$=$$Z_{\pi/2}X_{\pi/4}$ implements an IQOP for the basis X, that is, the input state $\vert 1_{D}\rangle$ becomes the state  $i\vert 1_{j'}\rangle$ at the output, and the state $\vert 1_{A}\rangle$ becomes the state  $i\vert 1_{j}\rangle$ at the output.
 \begin{figure}[h]
\centering
\includegraphics[width=8.5cm,clip]{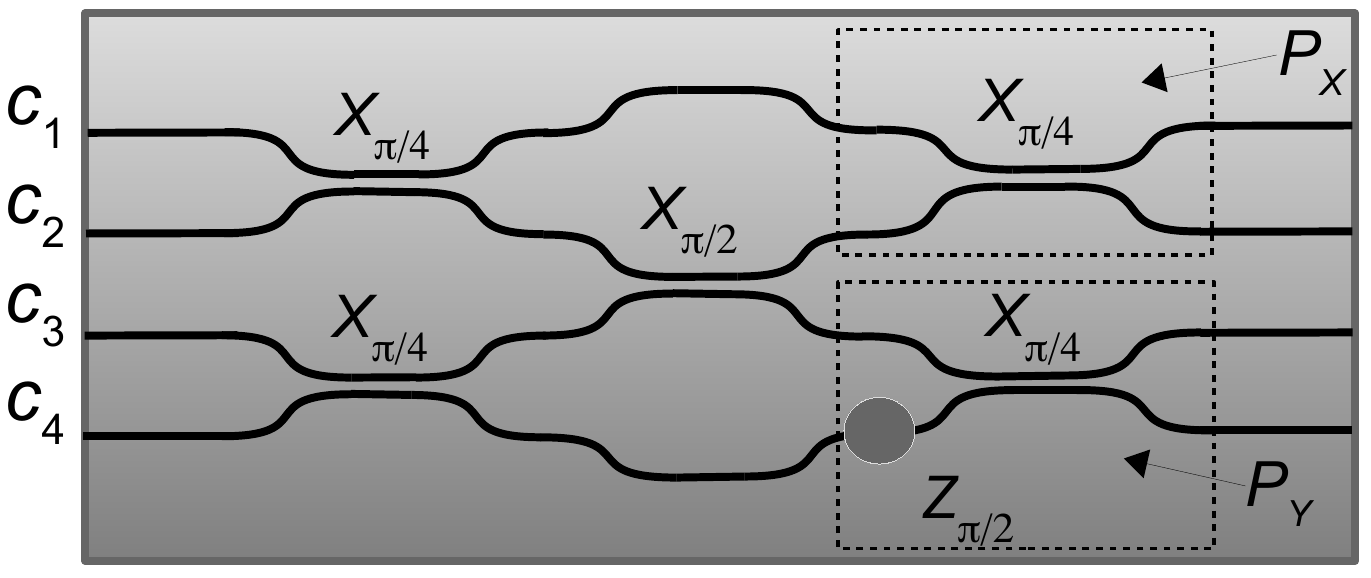}
\caption{Integrated quantum optical projector for  the two-dimensional bases X and Y.}
\label{QPP}      
\end{figure}
On the other hand,   a random choice of the measurement bases is required in quantum cryptography. For that, a more complex integrated circuit is needed. For example, in  Fig.~\ref{QPP} is shown an IQOC with four paths $c_{j},\,j=1,2,3,4$ for making projective measurements of states of bases X and Y in a random way.   Indeed, let us consider   input 1-qubits,  in a vector representation, $(c_{1},c_{3})$, coming from a transmission line where a  Quantum Key Distribution (QKD) is made, as for example the BB84 protocol (see, for instance, \cite{Gisin2002}). Likewise, let us consider that these 1-qubits are excited in  paths (optical modes) 1 and 3, that is,
 \begin{equation}
\vert L\rangle=c_{1}\vert 1_{1}\rangle+c_{3}\vert 1_{3}\rangle .
\end{equation}

 The first part of the IOC is formed by (see Fig.~\ref{QPP})  two couplers  ${\theta}=\pi/4$ along with a  coupler ${\theta}=\pi/2$ to implement a states division operation, that is
\begin{equation}
\mathbb{S} \vert L\rangle=\frac{1}{\sqrt{2}}\begin{pmatrix} 1&i&0&0 \\  0&0&i&-1\\ -1&i&0&0 \\ 0&0&i&1 \end{pmatrix}  \begin{pmatrix} c_{1}\\ 0 \\ c_{3}\\ 0 \end{pmatrix}=\frac{1}{\sqrt{2}}\begin{pmatrix} c_{1}\\ ic_{3} \\ -c_{1}\\ ic_{3}\end{pmatrix}.
\end{equation}
Note that, except phases, we have obtained a  state division, that is, a 1-ququart what makes sure a random choice of the measurement basis. The random 1-qubit states are $(c_{1},ic_{3})$ and $(c_{1},-ic_{3})$, therefore the $\pm \pi/2$ phases have to be taken into account in the projective measurement. Finally, the last couplers along with the phase shifter $Z_{\pi/2}$ allow to implement   the following projective transformation 
\begin{equation}
\mathbb{P}=\frac{1}{2}\begin{pmatrix} 1&i&-1&-i \\  i&-1&i&-1\\ -1&i&-i&-1 \\ -i&-1&-1&i \end{pmatrix}. 
\end{equation}
Indeed, the states of X basis, $\vert 1_{D}\rangle=(1/\sqrt{2})(\vert 1_{1}\rangle+\vert 1_{3}\rangle)$  and  $\vert 1_{A}\rangle=(1/\sqrt{2})(\vert 1_{1}\rangle-\vert 1_{3}\rangle)$, are measured  in a projective way at output channel guides 1 and 2, respectively, after propagation along  the coupler $X_{\pi/4}$.  Likewise,  the states $\vert 1_{L}\rangle=(1/\sqrt{2})(\vert 1_{1}\rangle+i\vert 1_{3}\rangle)$ and $\vert 1_{D}\rangle=(1/\sqrt{2})(\vert 1_{1}\rangle-i\vert 1_{3}\rangle)$ of the  Y basis are measured  at output channel guides 3 and 4, respectively, after propagation trough the phase shifter  $Z_{\pi/2}$  and the  coupler $X_{\pi/4}$. In short, the integrated component  works as a IQOP with random choice of measurement basis. Formally, we have a subproxector for states of the basis X with matrix $P_{X}=X_{\pi/4}Z_{\pi/2}$ (see Fig.~\ref{QPP}) and for the states of basis Y with the matrix $P_{Y}=-X_{\pi/4}$ (see Fig.~\ref{QPP}), that is, except an initial phase shifter $Z_{\pi/2}$  for X basis,  and a global phase $e^{i\pi}$ for  Y basis, the subprojectors are $X_{\pi/4}$  couplers. Therefore $X_{\pi/4}$ and $X_{\pi/2}$ can be regarded as  basic functional units of  IQOPs. Next, we demonstrate the fabrication of these units  in a IExG platform.  
 It is well known  that IExG platform has advantages with respect to cost, technological requirements and fibre optical compatibility, and moreover it could also  be  reconfigurable in a thermo-optical way. Finally, we will present the experimental results on optical characterization and optical tests of IQOPs for states of bases X and Y.

\section{Fabrication of a 2$\rm{x}$2 projector}
\label{sectionIII}

This section focus on the fabrication of the basic units  $X_{\pi/4}$ and $X_{\pi/2}$ given by 2x2 directional couplers  in glass by thermal ion-exchange processes in conjunction with a two-step photolithographic procedure.  These basic units allow to  implement IQOPs with arbitrary choice of measurement basis as shown in the above section.
Firstly, we briefly explain the details of the two consecutive purely thermal K$^{+}$/Na$^{+}$  diffusion process in order to modify the chemical composition of the substrate in a shallow layer just  a few micrometers deep. We must stress that this is a two-step process which is  different from the one proposed in reference \cite{Yip1984} where two consecutive K$^{+}$/Na$^{+}$ ion-exchanges are made, the first is selective and the second is planar. In our case, the first is also selective but the second one is a burial process. Secondly, we describe the photolithographic procedure along with all the optomechanical considerations in order to produce the final 2x2 IQOPs unit.

\begin{figure}[h]
\centering
\includegraphics[width=9.0cm,clip]{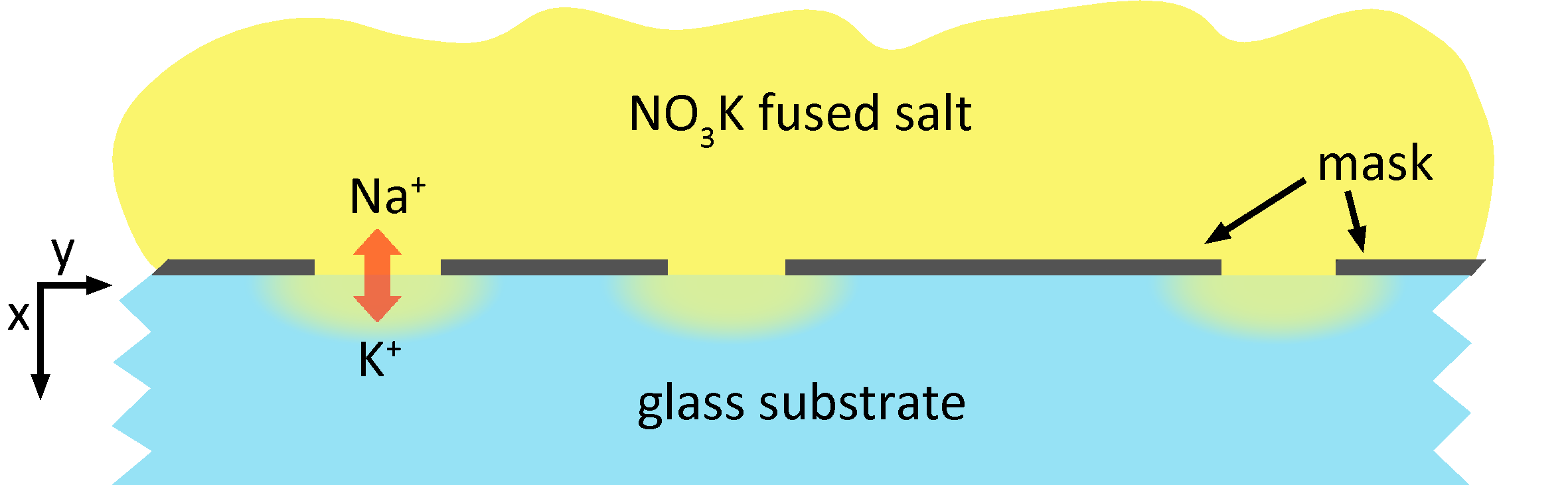}
\caption{\textcolor{black}{Sketch of the selective ion-exchange process. }}
\label{ScketchIExG}      
\end{figure}

Let us consider a a soda-lime glass substrate  dipped  in a salt of potassium nitrate (KNO3) at a high temperature $T$.  The Na$^{+}$ ions, inside the glass, acquire great mobility, then they come out of the glass and dissolve into the salt, and at the same time, K$^{+}$ ions of the salt can get inside the glass occupying the vacuum left by the Na$^{+}$ ions. The concentration of K$^{+}$ ions inside the glass, after the diffusion, results in an increase of the refractive index in a micrometrical film of the glass substrate. 
Thus, an integrated waveguide inside the glass is obtained. A subsequent thermal diffusion can be used once again to perform a burial process of this waveguide by immersing the substrate into a melted salt of NaNO3. The Na$^{+}$  ions present in the salt replace the most surface doping cations K$^{+}$, remaining those ones that are deeper inside the glass. The concentration of exchanged ions K$^{+}$ also expands inside the glass, which lowers the increase of refractive index \cite{Miliou:91}. This second step, i.e. the burial process, is not mandatory but presents two great advantages. By one hand,  the modal field is  located further  from the glass surface, which can present irregularities, and therefore losses are reduced \cite{Linares1994},  and on the other hand, it  reduces the typical anisotropy in Na$^{+}$/K$^{+}$ waveguides since the accumulated mechanical stresses, due to the difference in the size of the exchanged ions, are reduced \cite{Tervonen11}.  Moreover, if a selective exchange is required to produce  channel waveguides in the glass substrate, a mask with the suitable pattern of channel waveguides is required, for example, an aluminum  mask can be produced by some lithographic method with the IOC design imprinted in it, as shown, for example, in Fig.~\ref{ScketchIExG}. 
The first step in the manufacture of functional couplers is to get monomode channel waveguides. One dimension of the waveguide is related with the depth of the diffusion along $x$-direction, while the other one is related with the size along $y$-direction of the aperture of the metal mask and the lateral diffusion inherent to ion exchange processes. Fig.~\ref{ScketchIExG}  is a sketch of a selective ion-exchange process, where is shown  the mask, the fused salt and the change of refractive index   (concentration of exchanged ions) presenting an effective depth  and a lateral extension of the diffusion.  In order to select  appropriate parameters for the temperature ($T$) and diffusion time ($t_{d}$), we adopt the effective index method (EIM) \cite{Lee, Linares2000JMO, Linares2001JMO}. EIM will be applied to the case in which the waveguide only admits one mode in each dimension, and moreover in our case a separability approximation, justified by the weak guidance, is also considered. The condition for getting a single mode in depth was studied experimentally by  the optical characterization of planar waveguides. The condition in the parallel direction to the substrate ($y$-direction) was studied by numerical simulations based on the EIM.
As mentioned above, a two-steps photolithographic process is used to create a metal shield and select the areas where the ion exchange process will take place. We must indicate that first of all, the photolithographic process has  limiting condition since the optical system, developed in our lab to obtain a metallic mask on the surface of the substrate, has a resolution   that only allows to make channels wider than  3 microns. Second, the lateral diffusion also affects the size of the guide in that dimension (parallel to the surface, see Fig.~\ref{ScketchIExG}). This lateral diffusion extension  is estimated in relation to the depth of penetration of the ions and is determined from an experimental fit. 

On the other hand, the melting points of KNO3 and NaNO3 are about 334$^o$C and 307$^o$C respectively. From previous works  we have established a fixed temperature $T=400^o$C  for both the first ion-exchange and the second diffusion (burial process) in a soda-lime glass with index $n_{s}=1.5104$ for $\lambda_{o}=633$nm.   Several thermal diffusions in test glasses without metal masks were made and the effective index of the different modes of the waveguide were measured by prism coupling (also known by M-lines method), which was done with a automated system (Metricon, Model 2010/M). From this empiric calibration and taking into account numerical results by using the  EIM  method  we have  fixed  a diffusion time for the first diffusion  equal to $t_{d1}$$=$$30\,$min  and a time $t_{d2}$$=$$10\,$min  for the second  (burial) diffusion process whose primary aim, as commented above, was to reduce anisotropy.  By doing so,  
we obtain single-mode waveguides. It is worth indicating that  for  a wavelength $\lambda_{o}=633$nm the value  of the effective index of  the first planar ion-exchange is $N_{\text{\em eff}}=1.5125$ and the one of the  planar waveguide after the second ion exchange is  $N_{\text{\em eff}}=1.5117$. 

\begin{figure}[ht]
\centering
\includegraphics[width=1\linewidth]{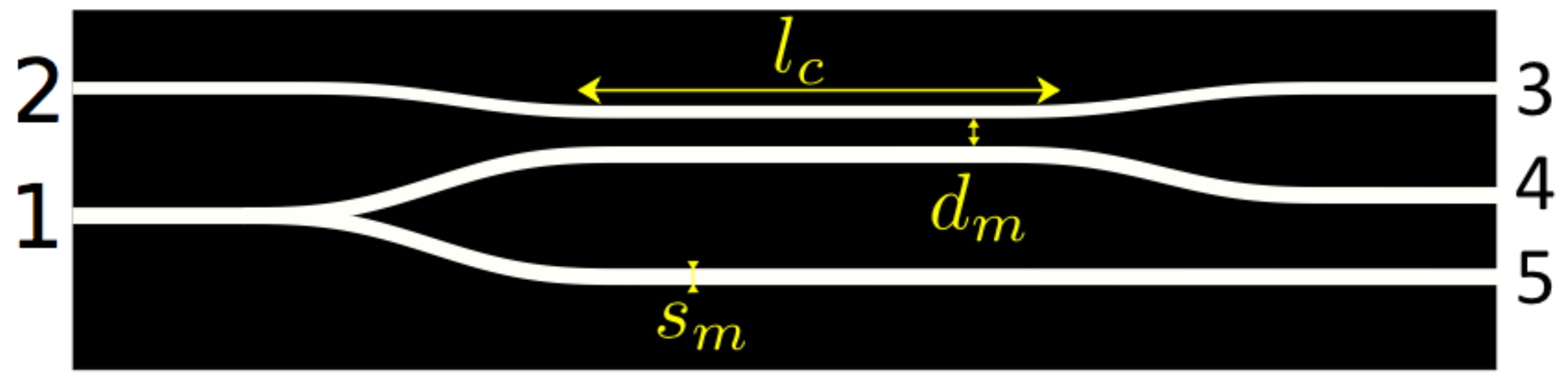}
\caption{Basic integrated optical element (BIOE) for optical characterization purposes.}
\label{basic_unit}
\end{figure}

In order to design and fabricate IQOPs a prior optical characterization of the full  fabrication process of basic elements as directional couplers as $X_{\pi/4}$, $X_{\pi/2}$ is needed. For that we will use a basic integrated optical element (BIOE) shown  in  Fig.~\ref{basic_unit}.  It consists of a set of three waveguides of width $s_{m}$ forming a Y union and a directional coupler which is object of optical characterization. The Y junction begins in the  initial waveguide 1  and has the  purpose of providing a reference signal \cite{Walker:83}.  Next,  the initial waveguide 2 and one of the  waveguides of Y junction are approached through curves following the Minford function \cite{Minford1982}, that is, a directional coupler is defined. The other channel waveguide (reference waveguide) of the Y junction follows unaltered to work as a reference. The light  coupled into the waveguide that goes to the Y junction is splitted 50\%-50\% between both waveguides. 
When the ion-exchange fabrication parameters are fixed (temperature, time, salt, ionic concentration and  glass substrate), then the spatial parameters $d_m$ (distance between coupled channel waveguides) and $l_c$ (coupling length) shown in  Fig.~\ref{basic_unit} will allow to find  the  particular parameters for  couplers as $X_{\pi/4}$, $X_{\pi/2}$ and  so on. 
To determine, for example, the parameters $d_m$ and $l_c$ for a particular coupler a fit method is applied, that is, the design of the mask includes a series of  BIOEs  with different $d_m$ and $l_c$ and with a double objective.  By one hand, it ensures that some of the BIOEs can become  equal  or quite close to, for example, the remarkable couplers $X_{\pi/4}$ or $X_{\pi/2}$, and  on the other hand  and importantly, allows to obtain the empiric curves of these parameters that in turn allow us to design the fabrication of  couplers with the desired coupling percentage in a second fabrication process.  For this purpose, we have  designed a mask containing 16 BIOEs in 4 blocks, each block maintains $d_m$ and $l_c$ is varied. The results of the optical characterization of the BIOEs and the empiric curves of the fabrication parameters are analyzed and shown in the next section. We must indicate that the  separations and the coupling lengths, after the photo-reduction process,  are found in the interval $d_{m}$\,$\in$\,$[3,7$.$5]\,\mu$m and  $l_{c}$\,$\in$\,$[0$.$5,2]\,$mm. We must also stress that the lateral diffusion increases the width of the channel guides up to 4$\mu$m with respect to the 3$\mu$m in the masks.
Finally, it must be indicated that these BIOEs were designed to introduce light through end-fire coupling, which consists of directly illuminate the extreme of the waveguide. Consequently, the waveguide have been ended in a flat and polished face, perpendicular to its propagation axis.

\section{Semiclassical demonstration of a 2$\rm{x}$2 projector}
\label{sectionIV}

\subsection{Optical characterization system}

\begin{figure*}[ht]
\centering
\includegraphics[width=0.9\linewidth]{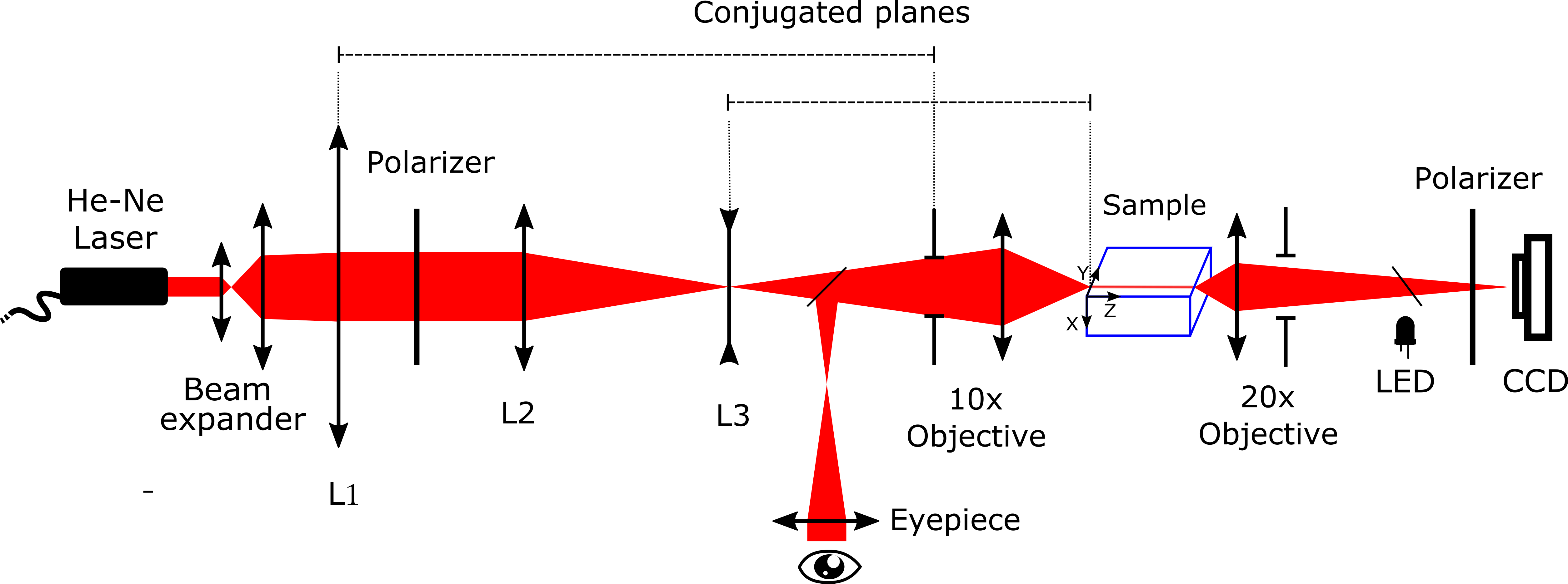}
\caption{Sketch of the optical setup for the characterization of the  BIOEs.}
\label{fig:optical_setup}
\end{figure*}

It is worth indicating that  channel waveguides of the BIOEs  were previously  inspected with a DIC transmission microscope  to verify that everything is correct and that the guides do not suffer from any blockage. Next,  to characterize their coupling performance, it is necessary to design and assemble an optical  system that allows to introduce light into channel waveguides through the end-fire coupling technique and at the same time analyze the light at the exit. In Fig.\ref{fig:optical_setup} a sketch  of said set-up is shown. This optical system  is similar to the one described in reference \cite{Prieto2012} with some small modifications.
First, an expanded He-Ne laser beam is refracted through a low-power positive lens L1, with focal $f=2$m,  that can be moved up and down to redirect the beam with a negligible change in its original divergence. Then, a high-power positive lens L2, with focal $f=0.1$m  is used to focus the previously expanded He-Ne laser beam. Next, the beam is focused again by means of a finite-conjugated (DIN standard) 10$\times$ microscope objective, placed at 160 mm of the image focal plane of L2 in order to focus a high-quality beam again. In this plane, an auxiliary negative lens L3 is placed to accomplish two main tasks: by one hand it helps to conjugate the plane of the low-power lens on the back focal plane of the 10x objective (usually close to the aperture stop, see Fig.~\ref{fig:optical_setup}), and on the other hand it facilitates the alignment by displacing it in the {\sc xy} plane allowing  to adjust the angle of incidence on the sample. 
The system allows  that the central ray of the beam is kept parallel to the optical axis, then, its phase is constant along the input face. To place just the input plane on the beam waist, the back-reflected beam from this face must maintain the same transversal size as the incident beam along the illumination system. Besides, to ensure that the beam initially falls on the waveguide, its output face is illuminated with a red LED;  
therefore, both the laser back-reflection (suitably attenuated) and the input face are simultaneously seen through the 10$\times$ microscope objective by using a cube beam-splitter (see Fig.~\ref{fig:optical_setup}) and an eyepiece. 
If the beam waist at the input face is fitted to the mode size, the beam matches the fundamental mode very well for a particular beam height. In this way, the best coupling coefficient to the first mode is achieved.  As the required waist is wider than the objective resolution, the Gaussian beam must illuminate only a portion of the objective aperture stop. In short, the illumination system generates a high-quality Gaussian beam with a waist of few micrometers wide on the input face of the waveguide to excite the fundamental mode of the waveguide. Moreover, the beam can be finely moved along the entrance face of the sample. At the output face, another standard 20$\times$ microscope objective produces an image of the end surface into a CMOS Thorlabs DCC1545M camera. Therefore, we can take images of the transmitted irradiance by the waveguides at the output face. Two polarizers were inserted in this set-up to work with an specific polarization (TE or TM). First polarizer ensures the polarization state at the entrance of the waveguide. Moreover, since ion exchange K$^{+}$/Na$^{+}$  in glass after burial process can still have a certain anisotropy, a second polarizer, placed in front of the CCD, allows us to distinguish between polarization states at the output.
\vspace{-0.35cm}


\subsection{Optical characterization results}
In order to characterize modal coupling in the BIOEs (16 units) fabricated in the glass substrate, a  He-Ne laser light is coupled to the  Y-junction of each BIOE of the substrate, in other words, a  coherent state $\alpha$ is used. Light is divided in the Y junction, with half of it going through the reference  waveguide 5 and the other half going through a waveguide (4) that is part of a coupler. Therefore, three  modal outputs 3, 4 and 5 are observed on the CMOS camera. 
\begin{figure}[ht]
\centering
\includegraphics[width=1\linewidth]{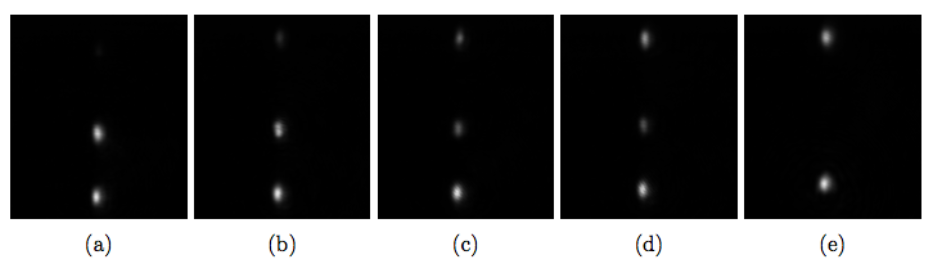}
\caption{Light at the output of different IQOPs obtained on a CMOS after introducing TE light into input 1 (Y union). The lower point corresponds to the output of the reference waveguide (5), while the two upper points are the outputs of the coupler waveguides (3 and 4).  (a) Light exits by output 4 ($\theta=\kappa z = 2 n \pi$). (e) Light exists by output 3 ($\theta = (2 n + 1) \pi/2$). (c) Light exits by both outputs 3 and 4 ($\theta = (2 n + 1) \pi/4$). (b) and (d) are intermediate cases.}
\label{fig:outputs}
\end{figure}
By taking pictures and processing them with a MATLAB software, we can evaluate the quantity of light that comes of each of the three waveguides, as well as the relative percentage of light that comes out in each of the waveguides of the coupler. The results are then compared by measuring the relative light intensity of the two outputs when light is coupled into input 2. This was done for the sixteen different BIOEs of the sample and for TE polarization. Qualitatively, it has been seen that the results are similar for TM. The presence of anisotropy in the guides makes a little difficult the measurements, detecting in some cases some energy transferred from TE to TM modes. In any case, the energy transferred was small and their study does not changes substantially the coupling results. In Fig.~\ref{fig:outputs} are presented  examples of  images taken during the optical  characterization procedure for the different BIOEs in the sample. 
\begin{table}[htbp]
	\centering
	\caption{Coupling results  expressed as normalized powers that exit from outputs  4 and 3, that is, ($P_{4}\vert P_{3}$).}	\begin{tabular}{|cc|cccc|}
		\hline
		\multicolumn{2}{|c}{\multirow{2}{*}{IOP Series}} & \multicolumn{4}{|c|}{$l_{c}$ (mm)}	\\
		\cline{3-6}
		& & 0.5 & 1.0 & 1.5 & 2.0 \\
		\hline
		\multicolumn{1}{|c|}{\multirow{4}{*}{$d_{m}$ ($\mu$m)}}  
		& 3.0 & (17.8$\vert$82.2) & (93.6$\vert$6.4) & (54.9$\vert$45.1) &  (\,2.5$\vert$97.5) \\
		\multicolumn{1}{|c|}{} 
		& 4.5 & (35.9$\vert$64.1) &  (\,5.9 $\vert$94.1) &  (\,7.4$\vert$92.6) & (40.9$\vert$59.1) \\
		\multicolumn{1}{|c|}{} 
		& 6.0 & (80.7$\vert$19.3) & (63.8$\vert$36.2) & (49.3$\vert$50.7) & (26.8$\vert$73.2) \\
		\multicolumn{1}{|c|}{} 
		& 7.5 & (92.6$\vert$\,7.4) & (88.4$\vert$11.6) & (86.3$\vert$13.7) & (79.8$\vert$20.2) \\
		
		\hline
	\end{tabular}
	\label{tab:AC2K2_Measurements}
\end{table}

\begin{figure}[ht]
\centering
\includegraphics[width=1\linewidth]{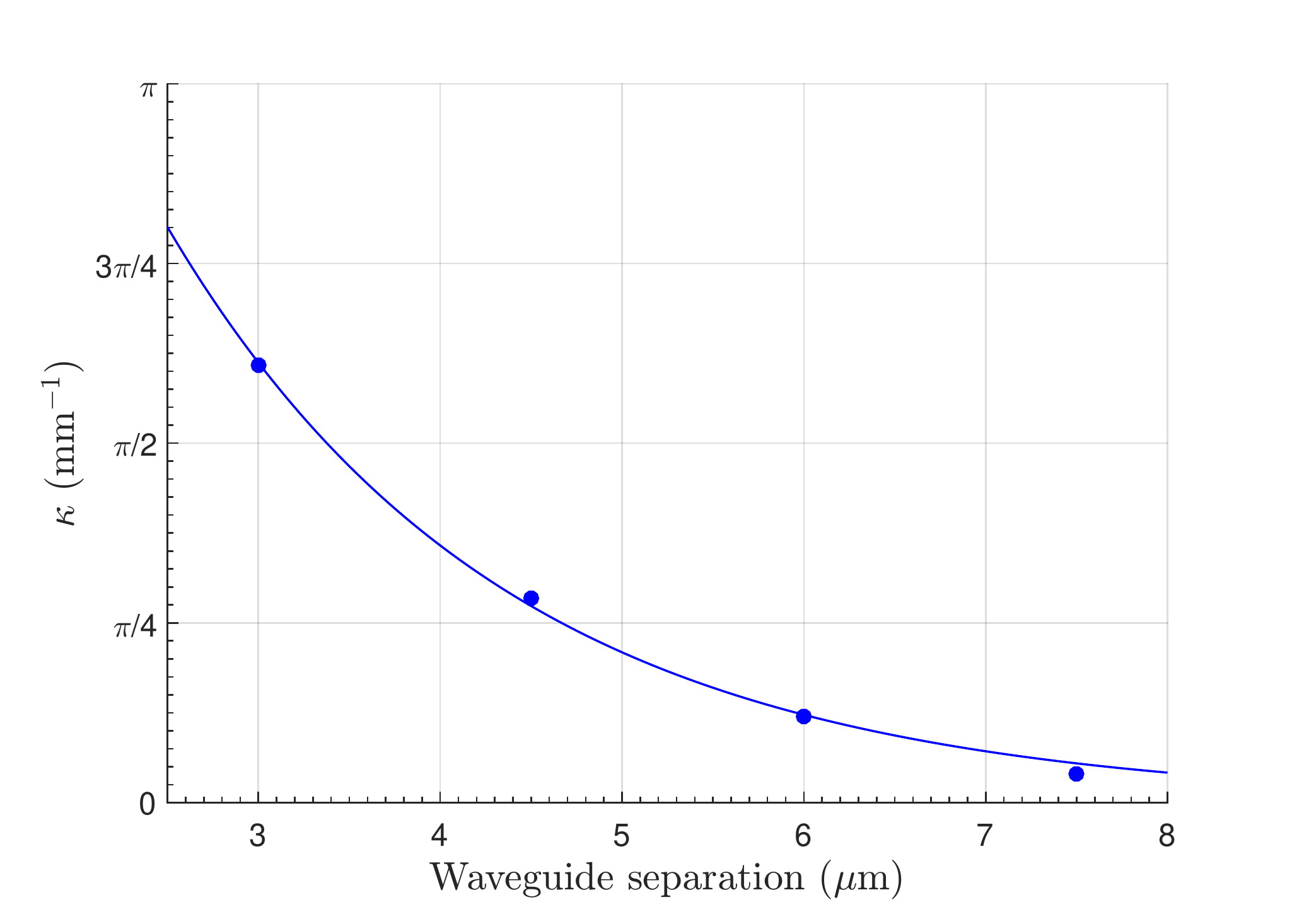}
\caption{Coupling constant $\kappa$ as a function of the waveguide separation $d_{m}$.}
\label{fig:fit3}
\end{figure}

Measurements of the relative intensities that comes out of outputs 4 and 3 are shown in Table~\ref{tab:AC2K2_Measurements}. According to Eq.~(\ref{Xcoupler}), these measurements correspond to $(P_{4}=\cos^{2}{\theta} \vert P_{3}=\sin^{2}{\theta})$. We must stress  that at Table~\ref{tab:AC2K2_Measurements} there are two BIOEs that work approximately as a $X_{{\pi}/{4}}$ coupler and three working approximately as a $X_{{\pi}/{2}}$ coupler, which are the most interesting couplers for this work. These cases, in $\left( d_{m}, l_{c} \right)$ notation, are $\left( 3.0, 1.5 \right)$ and $\left( 6.0, 1.5 \right)$  for $X_{{\pi}/{4}}$, and $\left( 4.5, 1.0 \right)$, $\left( 4.5, 1.5 \right)$ and $\left( 3.0, 2.0 \right)$  for $X_{{\pi}/{2}}$. Obviously, an optical characterization will provide empiric functions for determining the fabrication parameters in order to obtain quasi-exact couplers   $X_{{\pi}/{4}}$ and  $X_{{\pi}/{2}}$. 
For that, we can calculate the phase \smash{$\theta=\kappa (l_{c}+\Delta l_{c}) =\kappa L= \arccos{\sqrt{P_{4}}}$}, where $P_{4}$ is the relative power of output 4, where $L$ is an effective coupling length. Indeed, it is important to note that we  have introduced $\Delta l_{c}$ that represents an effective length  corresponding to  the interaction  before and after the region defined by $l_{c}$ in the masks. Alternatively, $L>l_c$ can be interpreted as the length of an effective coupler formed by two parallel channel waveguides. It must noted that  $\theta=\kappa (l_{c}+\Delta l_{c}) $ values obtained belong to the first quadrant, but have to be ordered to their right one.
 These phases  can be represented as a function of either coupler length $l_{c}$, that is, $\theta_{d_m}$, or waveguide separation $d_{m}$, that is, $\theta_{l_c}$. In the first case, as the coupling phase itself means, the relation between coupling phase and $l_{c}$ is linear. In the second case, we know that the coupling comes from the evanescent exponential fields  of the optical modes outside the waveguides. Thus, it is expected that the relation between the coupling phase and the waveguide separation would be exponential as well. Therefore, it should be possible to fit our measures to the following curves
	\begin{equation}
	\theta_{d_m}=\kappa (l_{c}+\Delta l_{c})= a_{l} \cdot l_{c} + b_{l},
	\label{eq:Coupling_Length_Fit}
	\end{equation}
	\begin{equation}
	\theta_{l_c}= 
	a_{e} \cdot  \exp \left( -b_{e} \cdot d_{m} \right).
	\label{eq:Coupling_Separation_Fit}
	\end{equation}
	\label{eq:Coupling_Characterization_Fit}
\begin{figure}[ht]
\centering
\includegraphics[width=1\linewidth]{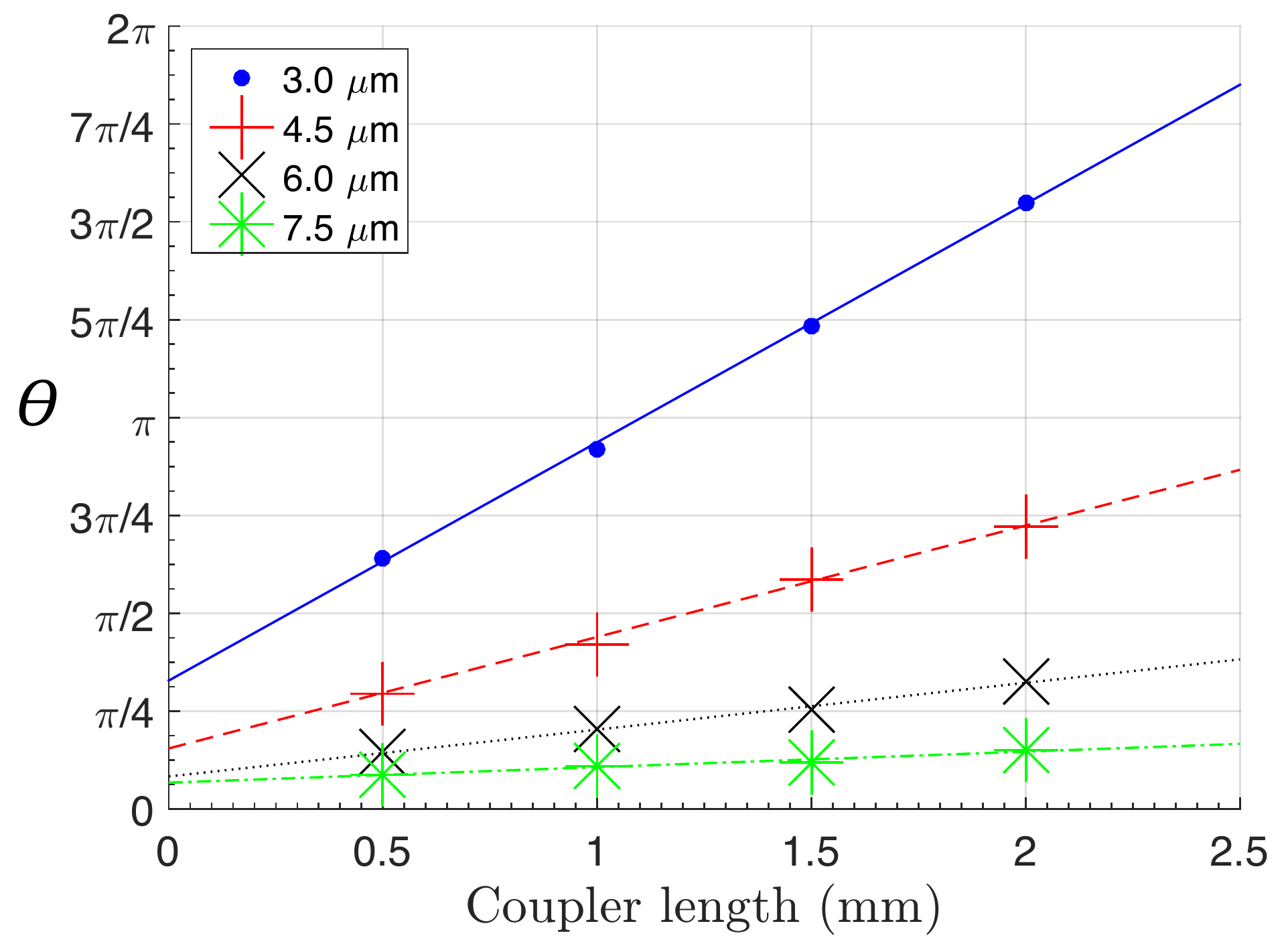}
\caption{Coupling phase $\theta$ as a function of the coupler length $l_{c}$ for different waveguides separations $d_{m}$.}
\label{fit1c}
\end{figure} 
\vspace{-0,35cm}
\ \\
where $a_{l}$, $b_{l}$, $a_{e}$ and $b_{e}$ are empirical fitting parameters.  Fig.~\ref{fit1c} shows the coupling phase $\theta$ as a linear  function of the coupler length $l_{c}$, according to Eq.~(\ref{eq:Coupling_Length_Fit}), for the different  values of the waveguide separations $d_{m}$.\begin{figure}[ht]
\centering
\includegraphics[width=1\linewidth]{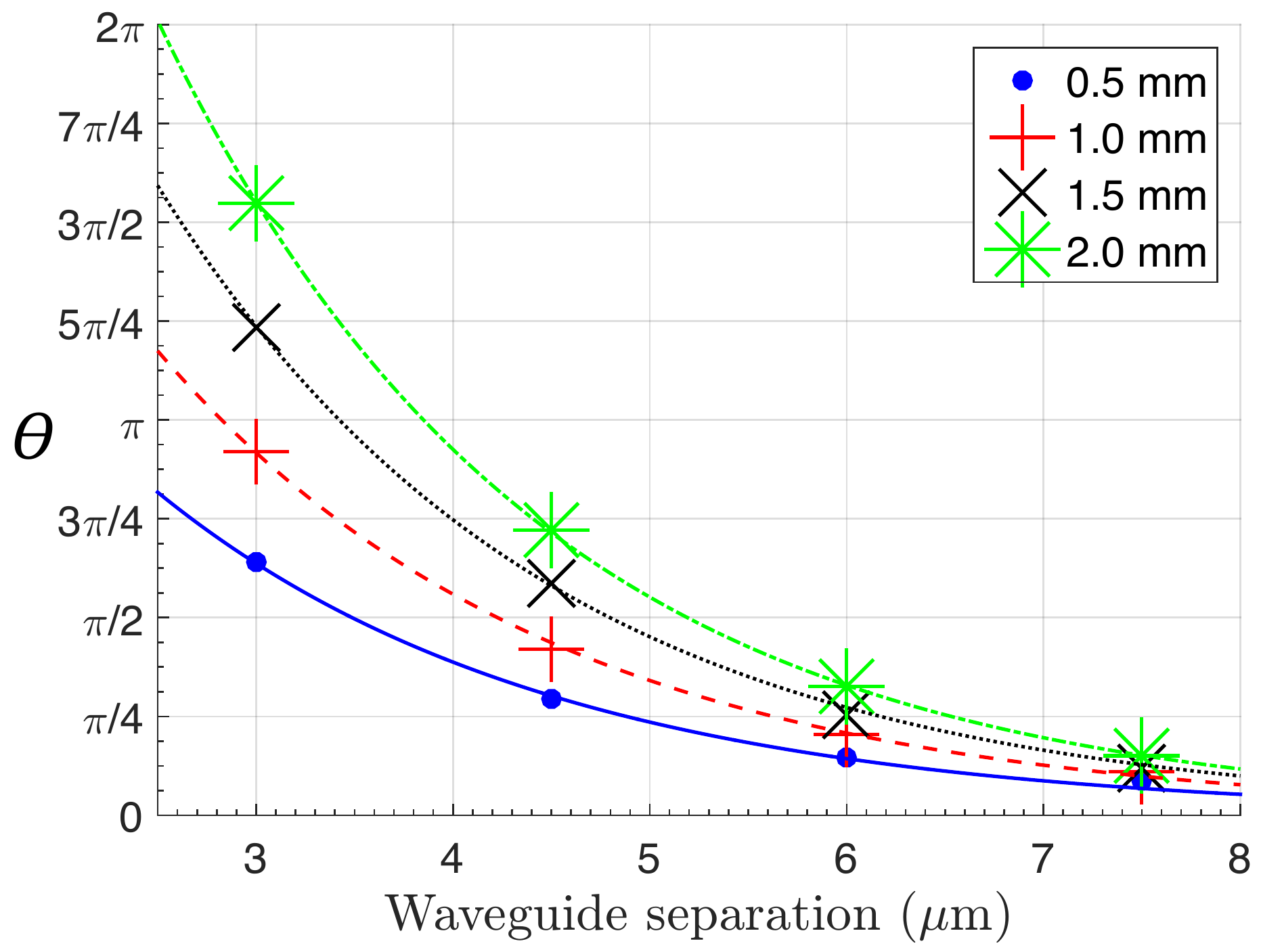}
\caption{Coupling  phase $\theta$ as a  function of the waveguide separation $d_m$ for different coupler lengths $l_{c}$.}
\label{fit2c}
\end{figure}
In Fig.~\ref{fit1c}, it is clearly seen how the coupling phase grows linearly with the coupler length. Besides, in the different curves we see that, the closer the waveguides are, the faster coupling phase grows, as the coupling is stronger.  On the other hand, Fig.\ref{fit2c} shows  curves of the coupling phase $\theta$ as a function of the waveguide separation,   according to Eq.~(\ref{eq:Coupling_Separation_Fit}),  for different  lengths $l_{c}$. An exponential decrease with the waveguide separation is clearly observed. 
These empirical fits provide remarkable  fabrication curves that allow us to manufacture any kind of coupler by Na$^{+}$/K$^{+}$ ion-exchange  with  waveguides of mask width of $s_{m} = 3$ $\mu$m. From the fitted slopes values  and collected from  Fig.~\ref{fit2c}  the dependence of the coupling constant $\kappa$ with the waveguide separation can be obtained by fitting them to  Eq~\ref{eq:Coupling_Separation_Fit}. The result of the fitting is shown in Fig.~\ref{fig:fit3} and corresponds to the following empiric function
\begin{equation} 
\begin{aligned}
 \kappa = 3.065 \pi \cdot   \exp \left[ -0.537 \cdot d_{m} \right] \, \text{\rm{mm}}^{-1}.
\end{aligned}
\label{eq:Coupling_Kappa_Fit}
\end{equation}
With these results, it is possible to fabricate different couplers $X_{\theta}$ by varying the separation and distance. Furthermore,  it has been fortunately verified that among the different BIOEs used for calibration, there is a almost perfect $X_{\pi/4}$ coupler, that is, an element that can act as an IQOP and therefore  allows us to carry out a projection test.

\subsection{Semiclassical test of the integrated projector}

The projection test consists of introducing  laser light   with different relative phases $\epsilon$  through  two input channel waveguides and observing the output  as a function of the relative phase. To avoid losses in the reference waveguide, the sample was turned around, or  in other words,  light is propagated from right to left in the BIOE shown in Fig.~\ref{basic_unit}. In this way, former outputs 3 and 4 are used now as the inputs where light is coupled, while the two original inputs 1 and 2 are now used as outputs, therefore it acts as a IQOP. In short, we illuminate the IQOP with a two-mode coherent state 
\begin{equation}
\vert L\rangle=\vert \alpha_{3}\alpha_{4}\rangle=\vert  a_{3}e^{i\epsilon_{3}} \,a_{4}e^{i\epsilon_{4}}\rangle
\end{equation}
with $a_{3,4}=\vert \alpha_{3,4}\vert$, where $\epsilon=\epsilon_{4}-\epsilon_{3}$ is the relative phase.  Nonetheless, some losses in output 1 due to the Y-junction were still observed. In this way, when light came out entirely from output 1, its intensity was lower than when it came exclusively from output 2.

\begin{figure}[ht]
\centering
\includegraphics[width=1\linewidth]{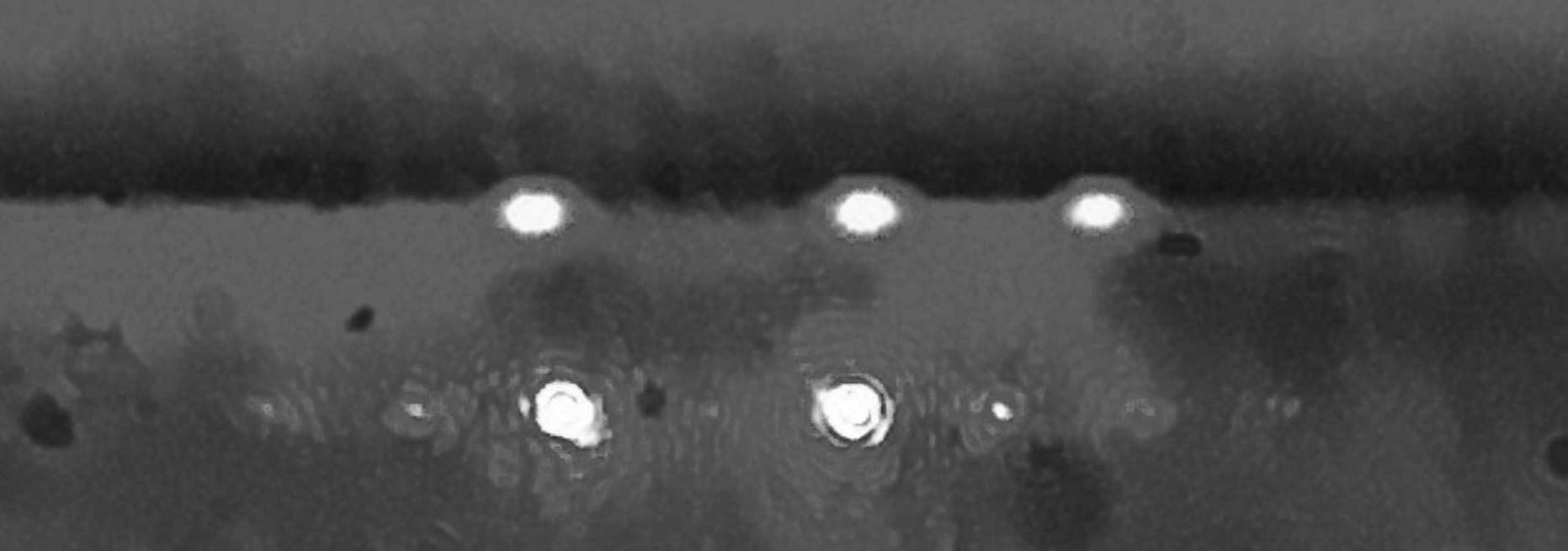}
\caption{Three channel waveguides of the IQOP (upper) and the two diffracted beams at their side (bottom), as observed through the ocular lens (note that optical saturation produces a slightly distorted image). }
\label{fig:picture_eyepiece}
\end{figure}

\begin{figure}[htbp]
		\centering
		\includegraphics[width=\linewidth]{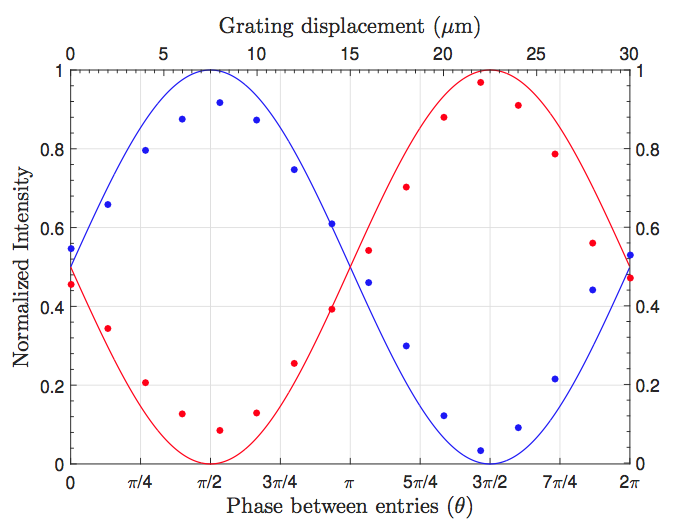}
		\caption{Light intensities in both outputs in function of the displacement of the grating (upper x-axis) and the introduced phase between beams (lower x-axis). The line represents the theoretical curve of a  $\pi/4$-coupler. Blue line represents output 1 and the red line, output 2.}
		\label{fig:Projection_Phase}
\end{figure}

In order to produce different relative  phases, we introduced a diffraction grating (DG) between the auxiliary lens L3 and the beam splitter indicated in Fig.~\ref{fig:optical_setup}. This DG has a period of 60 $\mu m$ and was designed to not generate zero order. Instead, light is mainly distributed symmetrically between diffraction orders +1 and -1, allowing us to obtain two light beams (coherent states) with the same intensity. This DF was also fabricated by  Na$^{+}$/Ag$^{+}$  ion exchange  and following the method described in reference \cite{Montero2020}. The DG is mounted in a support with two micrometrical screws, which allow us to move it in the {\sc x}  and {\sc z} axis. The two diffraction orders cross through the beam splitter and are focused in the sample by the objective. As made before, through the ocular we can observe these beams and thus to couple them into the waveguides, as seen in Figure \ref{fig:picture_eyepiece}. By displacing the DG closer or farther to the beam splitter, we can control the separation between focused beams to correctly couple each of them into the waveguides of the couplers. Moreover the same power $a_{3}=a_{4}=p_{o}$ is obtained in each diffraction order. On the other hand, by displacing transversally the DG ({\sc x} direction), a phase shift $\epsilon$ is added between the beams. Particularly, a displacement of half a period (30 $\mu m$) corresponds to a phase shift of $2\pi$ between them.
In short, the two-mode coherent states produced by the DG are given, in a good approximation, by  $\vert L\rangle = \vert p_{o} \,p_{o}e^{i\epsilon}\rangle$. Now, let us consider a very weak coherent state, that is, $p_{o}\ll1$, we obtain the single photon state  
\begin{equation}
\vert L\rangle \approx \vert 0\rangle+ p_{o}\vert 1_{3}\rangle + p_{o}e^{i\epsilon}\vert 1_{4}\rangle 
\end{equation}
Then, by taking into account the operator transformations of a $X_{\theta}$ coupler, the normalized detection probability of a photon coming out from any of outputs 1 and 2 of the coupler is given by 
\begin{equation}
\mathcal{P}_{1,2} =\frac{1}{2} (1 \pm \sin2\theta \,\sin\epsilon ),
\label{eq:Projection_Intensity}
\end{equation} 
which, except constants,  corresponds to the classical light power. In short,  a semiclassical analysis of the IQOP also provides a characterization for single photon states, that is, a test of the IQOP to analyze the capability for  implementing projective measurements can be made. 
We must recall that  couplers with $\theta = \pi/4$ implement  projections of states belonging to basis X and Y, that is, of the states   given by Eqs.~(\ref{baseX}) and (\ref{baseY}).  The  sample selected for testing is  the labeled by $(d_{m},l_{c})=(6.0, 1.5)$ in Table \ref{tab:AC2K2_Measurements}, that is, with $X_{\pi/4}$. By displacing the optical grating little by little  a few micrometers and taking measures of light powers in both outputs, we can characterize the coupling. Moreover, the measurements of powers at output 1 had to be corrected due to the losses in the Y-junction, by taking into account that these losses are given by  $(1-P_{1m}/P_{2m})$, where $P_{im}$ is the maximum measured power  when light comes out through outputs 1 and 2 (recall that the sample, that is, the coupler, was turned around). The results of the optical characterization of the coupler are represented in Fig.~\ref{fig:Projection_Phase} comparing it to theoretical curves given by Eq.~\ref{eq:Projection_Intensity}. It is observed the sinusoidal behavior of the light outputs as a  function of the grating displacement. Maximum and minimum experimental values do not reach 1 and 0 due to both the background light and the coupler which is not  actually a perfect $X_{\pi/4}$ coupler. We also observe that a full cycle is completed for a displacement of 30 $\mu$m, which corresponds to a phase $2\pi$.

The most remarkable cases are those ones corresponding to phases $\epsilon=\pi/2$, $3\pi/2$ as  these are the cases that simulate projective measurements of the quantum states belonging to base X and Y and commonly used  in the BB84 protocol.  Indeed,  the normalized input state without DG is given by  $(1/\sqrt{2})(\vert 1_{3}\rangle+\vert 1_{4}\rangle)$, then when DG  introduces phases   $\epsilon=\pi/2,\,3\pi/2$ the basis Y is implemented and the outputs states are  $i\vert 1_{1}\rangle$ and $\vert 1_{2}\rangle$, respectively, that is, the states have been projected on the outputs 1 and 2. In Fig.~\ref{fig:Projection_Phase} (b) and (c) we have the experimental results of these projections. Likewise, as analyzed in section \ref{sectionII}, if input states are $\vert 1_{L}\rangle$ and $\vert 1_{D}\rangle$, it requires the IQOP $Z_{\pi/2}X_{\pi/4}$. 
\begin{figure}[htbp]
	\centering
\includegraphics[width=\linewidth]{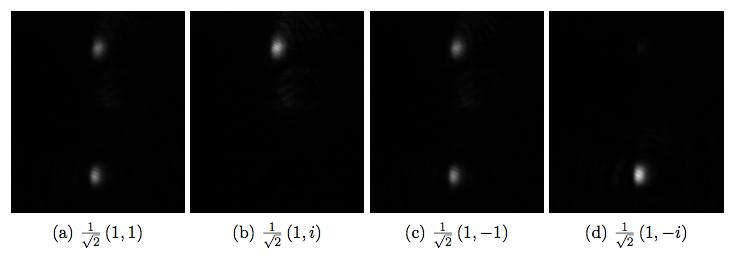}
	\caption{Output l states after introducing two identical light beams from both inputs changing the relative phase $\epsilon$ between them by displacing an optical grating.  (a) $\epsilon= 0$, (b) $\epsilon= \pi/2$, (c) $\epsilon= \pi$, (d) $\epsilon= 3\pi/2$.}
	\label{fig:Projection}
\end{figure}

We must stress that the above results with relative phases $\epsilon=\pi/2,\,3\pi/2$ introduced by the DG, can be reinterpreted as if the input state were $(1/\sqrt{2})(\vert 1_{3}\rangle\pm\vert 1_{4}\rangle)$ but with an additional phase $\pi/2$ introduced by the DG, that is, we simulate with the DG the phase shifter  $Z_{\pi/2}$. On the other hand,  we  see in  Fig.~\ref{fig:Projection_Phase},  for phase shifts $0$ and $\pi$,  that light comes out with almost same intensity in both outputs, that is, the states $\vert 1_{D}\rangle$ and $\vert 1_{A}\rangle$ are eigenstates of the IQOP. Therefore, the directional coupler works as an IQOP  for two dimensional quantum  states, in particular, for the states belonging to the basis Y, that is, $\vert 1_{L}\rangle=(1/\sqrt{2})(\vert 1_{3}\rangle+i\vert 1_{4}\rangle)$ and $\vert 1_{D}\rangle=(1/\sqrt{2})(\vert 1_{3}\rangle+i\vert 1_{4}\rangle)$. If an additional relative phase $\pi/2$ is introduced, that is,  a $Z_{\pi/2}$ phase shifter is added, then the coupler works as an IQOP    for  the two dimensional quantum  states $\vert 1_{D}\rangle$ and $\vert 1_{A}\rangle$ of basis X. We must stress that phase shifters can be implemented by changing the length of the channel guides, by changing the effective index in a small region and so on. In a future work, an optical characterization  of phase shifters fabricated by ion-exchange in glass will be presented.
\section{Conclusions}
\label{sectionV}
A series of basic integrated optical elements (directional couplers  plus Y junctions) have been  fabricated by ion-exchange Na$^{+}$/K$^{+}$ in soda-lime glass by using single mode channel waveguides fabricated with mask widths of 3$\mu$m and for 633\,nm. An optical characterization of these couplers has been made and empiric curves between fabrication parameters and optical parameters (as the coupling phase) have been obtained. In particular, coupling coefficient has been fitted to an exponential empiric curve respect to the separation of channel waveguides. These results allow to produce specific couplers like $X_{{\pi}/{4}}$ and  $X_{{\pi}/{2}}$ for fabricating IQOPs for quantum cryptography where a random choice of measurement basis is required. Likewise, the results can be used to produce quantum projectors of higher dimension  ($d>2$) or even to generate different quantum states, which is also an important task in quantum cryptography or in other devices for quantum information.  Likewise, an optical test of a $X_{{\pi}/{4}}$ coupler has been made by using an optical grating to generate  relative phases in single photon quantum states and thus to simulate quantum projections. The projections have been obtained with an efficiency close to 1.

\ \\
\textcolor{white}{sec}Authors wish to acknowledge the financial support of this work by Xunta de Galicia, Conseller\'ia de Educaci\'on, Universidades e FP, with a grant Consolidation-GRC Ref.-ED431C2018/11, a grant  Strategic Grouping of Materials (AeMAT)  Ref.-ED431E 2018/08, and a predoctoral grant (D. Balado, 2017), co-financied with the European Social Fund.


\section*{References}


\bibliography{PreprintProj.bib}

\begin{thebibliography}{27}%
\makeatletter
\providecommand \@ifxundefined [1]{%
 \@ifx{#1\undefined}
}%
\providecommand \@ifnum [1]{%
 \ifnum #1\expandafter \@firstoftwo
 \else \expandafter \@secondoftwo
 \fi
}%
\providecommand \@ifx [1]{%
 \ifx #1\expandafter \@firstoftwo
 \else \expandafter \@secondoftwo
 \fi
}%
\providecommand \natexlab [1]{#1}%
\providecommand \enquote  [1]{``#1''}%
\providecommand \bibnamefont  [1]{#1}%
\providecommand \bibfnamefont [1]{#1}%
\providecommand \citenamefont [1]{#1}%
\providecommand \href@noop [0]{\@secondoftwo}%
\providecommand \href [0]{\begingroup \@sanitize@url \@href}%
\providecommand \@href[1]{\@@startlink{#1}\@@href}%
\providecommand \@@href[1]{\endgroup#1\@@endlink}%
\providecommand \@sanitize@url [0]{\catcode `\\12\catcode `\$12\catcode
  `\&12\catcode `\#12\catcode `\^12\catcode `\_12\catcode `\%12\relax}%
\providecommand \@@startlink[1]{}%
\providecommand \@@endlink[0]{}%
\providecommand \url  [0]{\begingroup\@sanitize@url \@url }%
\providecommand \@url [1]{\endgroup\@href {#1}{\urlprefix }}%
\providecommand \urlprefix  [0]{URL }%
\providecommand \Eprint [0]{\href }%
\providecommand \doibase [0]{http://dx.doi.org/}%
\providecommand \selectlanguage [0]{\@gobble}%
\providecommand \bibinfo  [0]{\@secondoftwo}%
\providecommand \bibfield  [0]{\@secondoftwo}%
\providecommand \translation [1]{[#1]}%
\providecommand \BibitemOpen [0]{}%
\providecommand \bibitemStop [0]{}%
\providecommand \bibitemNoStop [0]{.\EOS\space}%
\providecommand \EOS [0]{\spacefactor3000\relax}%
\providecommand \BibitemShut  [1]{\csname bibitem#1\endcsname}%
\let\auto@bib@innerbib\@empty
\bibitem [{\citenamefont {Fox}(2006)}]{Fox}%
  \BibitemOpen
  \bibfield  {author} {\bibinfo {author} {\bibfnamefont {M.}~\bibnamefont
  {Fox}},\ }\href@noop {} {\emph {\bibinfo {title} {Quantum Optics}}}\
  (\bibinfo  {publisher} {Oxford Unversity Press},\ \bibinfo {year}
  {2006})\BibitemShut {NoStop}%
\bibitem [{\citenamefont {Agarwal}(2013)}]{AgarwalBook}%
  \BibitemOpen
  \bibfield  {author} {\bibinfo {author} {\bibfnamefont {G.}~\bibnamefont
  {Agarwal}},\ }\href@noop {} {\emph {\bibinfo {title} {Quantum Optics}}}\
  (\bibinfo  {publisher} {Cambridge Unversity Press},\ \bibinfo {address}
  {Cambridge},\ \bibinfo {year} {2013})\BibitemShut {NoStop}%
\bibitem [{\citenamefont {Hayashi}\ and\ \citenamefont
  {Nakanishi}(2018)}]{Overview}%
  \BibitemOpen
  \bibfield  {author} {\bibinfo {author} {\bibfnamefont {T.}~\bibnamefont
  {Hayashi}}\ and\ \bibinfo {author} {\bibfnamefont {T.}~\bibnamefont
  {Nakanishi}},\ }\href@noop {} {\bibfield  {journal} {\bibinfo  {journal} {SEI
  Technical Review}\ }\textbf {\bibinfo {volume} {86}},\ \bibinfo {pages} {23}
  (\bibinfo {year} {2018})}\BibitemShut {NoStop}%
\bibitem [{\citenamefont {Ca\~nas}\ \emph {et~al.}(2017)\citenamefont
  {Ca\~nas}, \citenamefont {Vera}, \citenamefont {Cari\~ne}, \citenamefont
  {Gonz\'alez}, \citenamefont {Cardenas}, \citenamefont {Connolly},
  \citenamefont {Przysiezna}, \citenamefont {G\'omez}, \citenamefont
  {Figueroa}, \citenamefont {Vallone}, \citenamefont {Villoresi}, \citenamefont
  {da~Silva}, \citenamefont {Xavier},\ and\ \citenamefont {Lima}}]{Canas2017}%
  \BibitemOpen
  \bibfield  {author} {\bibinfo {author} {\bibfnamefont {G.}~\bibnamefont
  {Ca\~nas}}, \bibinfo {author} {\bibfnamefont {N.}~\bibnamefont {Vera}},
  \bibinfo {author} {\bibfnamefont {J.}~\bibnamefont {Cari\~ne}}, \bibinfo
  {author} {\bibfnamefont {P.}~\bibnamefont {Gonz\'alez}}, \bibinfo {author}
  {\bibfnamefont {J.}~\bibnamefont {Cardenas}}, \bibinfo {author}
  {\bibfnamefont {P.~W.~R.}\ \bibnamefont {Connolly}}, \bibinfo {author}
  {\bibfnamefont {A.}~\bibnamefont {Przysiezna}}, \bibinfo {author}
  {\bibfnamefont {E.~S.}\ \bibnamefont {G\'omez}}, \bibinfo {author}
  {\bibfnamefont {M.}~\bibnamefont {Figueroa}}, \bibinfo {author}
  {\bibfnamefont {G.}~\bibnamefont {Vallone}}, \bibinfo {author} {\bibfnamefont
  {P.}~\bibnamefont {Villoresi}}, \bibinfo {author} {\bibfnamefont {T.~F.}\
  \bibnamefont {da~Silva}}, \bibinfo {author} {\bibfnamefont {G.~B.}\
  \bibnamefont {Xavier}}, \ and\ \bibinfo {author} {\bibfnamefont
  {G.}~\bibnamefont {Lima}},\ }\href
  {https://link.aps.org/doi/10.1103/PhysRevA.96.022317} {\bibfield  {journal}
  {\bibinfo  {journal} {Phys. Rev. A}\ }\textbf {\bibinfo {volume} {96}},\
  \bibinfo {pages} {022317} (\bibinfo {year} {2017})}\BibitemShut {NoStop}%
\bibitem [{\citenamefont {Balado}\ \emph {et~al.}(2019)\citenamefont {Balado},
  \citenamefont {Li$\tilde{\rm n}$ares},\ and\ \citenamefont {Xes\'us
  Prieto-Blanco}}]{Balado2019}%
  \BibitemOpen
  \bibfield  {author} {\bibinfo {author} {\bibfnamefont {D.}~\bibnamefont
  {Balado}}, \bibinfo {author} {\bibfnamefont {J.}~\bibnamefont {Li$\tilde{\rm
  n}$ares}}, \ and\ \bibinfo {author} {\bibfnamefont {D.~B.}\ \bibnamefont
  {Xes\'us Prieto-Blanco}},\ }\href@noop {} {\bibfield  {journal} {\bibinfo
  {journal} {JOSA B}\ }\textbf {\bibinfo {volume} {36}},\ \bibinfo {pages}
  {2793} (\bibinfo {year} {2019})}\BibitemShut {NoStop}%
\bibitem [{\citenamefont {Bai}\ \emph {et~al.}(2012)\citenamefont {Bai},
  \citenamefont {Ip}, \citenamefont {Huang}, \citenamefont {Mateo},
  \citenamefont {Yaman}, \citenamefont {Li}, \citenamefont {Bickham},
  \citenamefont {Ten}, \citenamefont {Li$\tilde{\rm n}$ares}, , \citenamefont
  {Montero}, \citenamefont {Moreno}, \citenamefont {Prieto}, \citenamefont
  {Tse}, \citenamefont {Chung}, \citenamefont {Lau}, \citenamefont {Tam},
  \citenamefont {Lu}, \citenamefont {Luo}, \citenamefont {Peng}, \citenamefont
  {Li},\ and\ \citenamefont {Wang}}]{Bai12}%
  \BibitemOpen
  \bibfield  {author} {\bibinfo {author} {\bibfnamefont {N.}~\bibnamefont
  {Bai}}, \bibinfo {author} {\bibfnamefont {E.}~\bibnamefont {Ip}}, \bibinfo
  {author} {\bibfnamefont {Y.-K.}\ \bibnamefont {Huang}}, \bibinfo {author}
  {\bibfnamefont {E.}~\bibnamefont {Mateo}}, \bibinfo {author} {\bibfnamefont
  {F.}~\bibnamefont {Yaman}}, \bibinfo {author} {\bibfnamefont {M.-J.}\
  \bibnamefont {Li}}, \bibinfo {author} {\bibfnamefont {S.}~\bibnamefont
  {Bickham}}, \bibinfo {author} {\bibfnamefont {S.}~\bibnamefont {Ten}},
  \bibinfo {author} {\bibfnamefont {J.}~\bibnamefont {Li$\tilde{\rm n}$ares}},
  , \bibinfo {author} {\bibfnamefont {C.}~\bibnamefont {Montero}}, \bibinfo
  {author} {\bibfnamefont {V.}~\bibnamefont {Moreno}}, \bibinfo {author}
  {\bibfnamefont {X.}~\bibnamefont {Prieto}}, \bibinfo {author} {\bibfnamefont
  {V.}~\bibnamefont {Tse}}, \bibinfo {author} {\bibfnamefont {K.~M.}\
  \bibnamefont {Chung}}, \bibinfo {author} {\bibfnamefont {A.~P.~T.}\
  \bibnamefont {Lau}}, \bibinfo {author} {\bibfnamefont {H.-Y.}\ \bibnamefont
  {Tam}}, \bibinfo {author} {\bibfnamefont {C.}~\bibnamefont {Lu}}, \bibinfo
  {author} {\bibfnamefont {Y.}~\bibnamefont {Luo}}, \bibinfo {author}
  {\bibfnamefont {G.-D.}\ \bibnamefont {Peng}}, \bibinfo {author}
  {\bibfnamefont {G.}~\bibnamefont {Li}}, \ and\ \bibinfo {author}
  {\bibfnamefont {T.}~\bibnamefont {Wang}},\ }\href
  {http://www.opticsexpress.org/abstract.cfm?URI=oe-20-3-2668} {\bibfield
  {journal} {\bibinfo  {journal} {Opt. Express}\ }\textbf {\bibinfo {volume}
  {20}},\ \bibinfo {pages} {2668} (\bibinfo {year} {2012})}\BibitemShut
  {NoStop}%
\bibitem [{\citenamefont {Balado-Souto}\ \emph {et~al.}(2019)\citenamefont
  {Balado-Souto}, \citenamefont {Li$\tilde{\rm n}$ares},\ and\ \citenamefont
  {Prieto-Blanco}}]{BaladoJMO}%
  \BibitemOpen
  \bibfield  {author} {\bibinfo {author} {\bibfnamefont {D.}~\bibnamefont
  {Balado-Souto}}, \bibinfo {author} {\bibfnamefont {J.}~\bibnamefont
  {Li$\tilde{\rm n}$ares}}, \ and\ \bibinfo {author} {\bibfnamefont
  {X.}~\bibnamefont {Prieto-Blanco}},\ }\href@noop {} {\bibfield  {journal}
  {\bibinfo  {journal} {J.Mod.Opt.}\ }\textbf {\bibinfo {volume} {66}},\
  \bibinfo {pages} {947} (\bibinfo {year} {2019})}\BibitemShut {NoStop}%
\bibitem [{\citenamefont {Leon-Saval}\ \emph {et~al.}(2013)\citenamefont
  {Leon-Saval}, \citenamefont {Argyros},\ and\ \citenamefont
  {Bland-Hawthorn}}]{Leon2013}%
  \BibitemOpen
  \bibfield  {author} {\bibinfo {author} {\bibfnamefont {S.}~\bibnamefont
  {Leon-Saval}}, \bibinfo {author} {\bibfnamefont {A.}~\bibnamefont {Argyros}},
  \ and\ \bibinfo {author} {\bibfnamefont {J.}~\bibnamefont {Bland-Hawthorn}},\
  }\href@noop {} {\bibfield  {journal} {\bibinfo  {journal} {Nanophotonics}\
  }\textbf {\bibinfo {volume} {2}},\ \bibinfo {pages} {429} (\bibinfo {year}
  {2013})}\BibitemShut {NoStop}%
\bibitem [{\citenamefont {Lee}(1986)}]{Lee}%
  \BibitemOpen
  \bibfield  {author} {\bibinfo {author} {\bibfnamefont {D.}~\bibnamefont
  {Lee}},\ }\href@noop {} {\emph {\bibinfo {title} {Electromagnetic Principles
  of Integrated Optics}}}\ (\bibinfo  {publisher} {Wiley, New York},\ \bibinfo
  {year} {1986})\BibitemShut {NoStop}%
\bibitem [{\citenamefont {Najafi}(1986)}]{Najafi92}%
  \BibitemOpen
  \bibfield  {author} {\bibinfo {author} {\bibfnamefont {S.~I.}\ \bibnamefont
  {Najafi}},\ }\href@noop {} {\emph {\bibinfo {title} {Introduction to Glass
  Integrated Optics}}}\ (\bibinfo  {publisher} {Artech House, Boston},\
  \bibinfo {year} {1986})\BibitemShut {NoStop}%
\bibitem [{\citenamefont {Li\~nares}\ \emph {et~al.}(2000)\citenamefont
  {Li\~nares}, \citenamefont {Montero}, \citenamefont {Moreno}, \citenamefont
  {Nistal}, \citenamefont {Prieto}, \citenamefont {Salgueiro},\ and\
  \citenamefont {Sotelo}}]{SPIE2000}%
  \BibitemOpen
  \bibfield  {author} {\bibinfo {author} {\bibfnamefont {J.}~\bibnamefont
  {Li\~nares}}, \bibinfo {author} {\bibfnamefont {C.}~\bibnamefont {Montero}},
  \bibinfo {author} {\bibfnamefont {V.}~\bibnamefont {Moreno}}, \bibinfo
  {author} {\bibfnamefont {M.~C.}\ \bibnamefont {Nistal}}, \bibinfo {author}
  {\bibfnamefont {X.}~\bibnamefont {Prieto}}, \bibinfo {author} {\bibfnamefont
  {J.}~\bibnamefont {Salgueiro}}, \ and\ \bibinfo {author} {\bibfnamefont
  {D.}~\bibnamefont {Sotelo}},\ }\href@noop {} {\bibfield  {journal} {\bibinfo
  {journal} {Proc. SPIE, Integrated Optics Devices IV}\ }\textbf {\bibinfo
  {volume} {3936}},\ \bibinfo {pages} {227} (\bibinfo {year}
  {2000})}\BibitemShut {NoStop}%
\bibitem [{\citenamefont {Righini}\ \emph {et~al.}(1997)\citenamefont
  {Righini}, \citenamefont {Conti},\ and\ \citenamefont
  {Forastiere}}]{Righini1997}%
  \BibitemOpen
  \bibfield  {author} {\bibinfo {author} {\bibfnamefont {G.~C.}\ \bibnamefont
  {Righini}}, \bibinfo {author} {\bibfnamefont {G.~N.}\ \bibnamefont {Conti}},
  \ and\ \bibinfo {author} {\bibfnamefont {M.~A.}\ \bibnamefont {Forastiere}},\
  }\href@noop {} {\bibfield  {journal} {\bibinfo  {journal} {Proceedings of
  SPIE}\ }\textbf {\bibinfo {volume} {2997}},\ \bibinfo {pages} {212} (\bibinfo
  {year} {1997})}\BibitemShut {NoStop}%
\bibitem [{\citenamefont {{Politi}}\ \emph {et~al.}(2009)\citenamefont
  {{Politi}}, \citenamefont {{Matthews}}, \citenamefont {{Thompson}},\ and\
  \citenamefont {{O'Brien}}}]{Politi2009}%
  \BibitemOpen
  \bibfield  {author} {\bibinfo {author} {\bibfnamefont {A.}~\bibnamefont
  {{Politi}}}, \bibinfo {author} {\bibfnamefont {J.~C.~F.}\ \bibnamefont
  {{Matthews}}}, \bibinfo {author} {\bibfnamefont {M.~G.}\ \bibnamefont
  {{Thompson}}}, \ and\ \bibinfo {author} {\bibfnamefont {J.~L.}\ \bibnamefont
  {{O'Brien}}},\ }\href {\doibase 10.1109/JSTQE.2009.2026060} {\bibfield
  {journal} {\bibinfo  {journal} {IEEE Journal of Selected Topics in Quantum
  Electronics}\ }\textbf {\bibinfo {volume} {15}},\ \bibinfo {pages} {1673}
  (\bibinfo {year} {2009})}\BibitemShut {NoStop}%
\bibitem [{\citenamefont {Wang}\ \emph {et~al.}(2020)\citenamefont {Wang},
  \citenamefont {Sciarrinom F.~Laing},\ and\ \citenamefont
  {Thompson}}]{Wang2020}%
  \BibitemOpen
  \bibfield  {author} {\bibinfo {author} {\bibfnamefont {J.}~\bibnamefont
  {Wang}}, \bibinfo {author} {\bibfnamefont {A.}~\bibnamefont {Sciarrinom
  F.~Laing}}, \ and\ \bibinfo {author} {\bibfnamefont {M.}~\bibnamefont
  {Thompson}},\ }\href {\doibase 10.1109/JSTQE.2009.2026060} {\bibfield
  {journal} {\bibinfo  {journal} {Nature Photonics}\ }\textbf {\bibinfo
  {volume} {14}},\ \bibinfo {pages} {273} (\bibinfo {year} {2020})}\BibitemShut
  {NoStop}%
\bibitem [{\citenamefont {Li$\tilde{\rm n}$ares}\ \emph
  {et~al.}(2011)\citenamefont {Li$\tilde{\rm n}$ares}, \citenamefont {Nistal},
  \citenamefont {Barral}, \citenamefont {Moreno}, \citenamefont {Montero},\
  and\ \citenamefont {Prieto}}]{Linares2011}%
  \BibitemOpen
  \bibfield  {author} {\bibinfo {author} {\bibfnamefont {J.}~\bibnamefont
  {Li$\tilde{\rm n}$ares}}, \bibinfo {author} {\bibfnamefont {M.~C.}\
  \bibnamefont {Nistal}}, \bibinfo {author} {\bibfnamefont {D.}~\bibnamefont
  {Barral}}, \bibinfo {author} {\bibfnamefont {V.}~\bibnamefont {Moreno}},
  \bibinfo {author} {\bibfnamefont {C.}~\bibnamefont {Montero}}, \ and\
  \bibinfo {author} {\bibfnamefont {X.}~\bibnamefont {Prieto}},\ }\href@noop {}
  {\bibfield  {journal} {\bibinfo  {journal} {Opt. Pura Apl.}\ }\textbf
  {\bibinfo {volume} {44}},\ \bibinfo {pages} {241} (\bibinfo {year}
  {2011})}\BibitemShut {NoStop}%
\bibitem [{\citenamefont {Spring}\ \emph {et~al.}(2013)\citenamefont {Spring},
  \citenamefont {Metcalf}, \citenamefont {Humphreys}, \citenamefont
  {Kolthammer}, \citenamefont {Jin}, \citenamefont {Barbieri}, \citenamefont
  {Datta}, \citenamefont {Thomas-Peter}, \citenamefont {Langford},
  \citenamefont {Kundys}, \citenamefont {Gates}, \citenamefont {Smith},
  \citenamefont {Smith},\ and\ \citenamefont {A.~Walmsley}}]{boson}%
  \BibitemOpen
  \bibfield  {author} {\bibinfo {author} {\bibfnamefont {J.}~\bibnamefont
  {Spring}}, \bibinfo {author} {\bibfnamefont {B.}~\bibnamefont {Metcalf}},
  \bibinfo {author} {\bibfnamefont {P.}~\bibnamefont {Humphreys}}, \bibinfo
  {author} {\bibfnamefont {W.}~\bibnamefont {Kolthammer}}, \bibinfo {author}
  {\bibfnamefont {X.}~\bibnamefont {Jin}}, \bibinfo {author} {\bibfnamefont
  {M.}~\bibnamefont {Barbieri}}, \bibinfo {author} {\bibfnamefont
  {A.}~\bibnamefont {Datta}}, \bibinfo {author} {\bibfnamefont
  {N.}~\bibnamefont {Thomas-Peter}}, \bibinfo {author} {\bibfnamefont {N.~K.}\
  \bibnamefont {Langford}}, \bibinfo {author} {\bibfnamefont {D.}~\bibnamefont
  {Kundys}}, \bibinfo {author} {\bibfnamefont {J.~C.}\ \bibnamefont {Gates}},
  \bibinfo {author} {\bibfnamefont {J.~B.}\ \bibnamefont {Smith}}, \bibinfo
  {author} {\bibfnamefont {P.}~\bibnamefont {Smith}}, \ and\ \bibinfo {author}
  {\bibfnamefont {I.~A.}\ \bibnamefont {A.~Walmsley}},\ }\href@noop {}
  {\bibfield  {journal} {\bibinfo  {journal} {Science}\ }\textbf {\bibinfo
  {volume} {239}},\ \bibinfo {pages} {798} (\bibinfo {year}
  {2013})}\BibitemShut {NoStop}%
\bibitem [{\citenamefont {Gisin}\ \emph {et~al.}(2002)\citenamefont {Gisin},
  \citenamefont {Ribordy}, \citenamefont {Tittel},\ and\ \citenamefont
  {Zbinden}}]{Gisin2002}%
  \BibitemOpen
  \bibfield  {author} {\bibinfo {author} {\bibfnamefont {N.}~\bibnamefont
  {Gisin}}, \bibinfo {author} {\bibfnamefont {G.}~\bibnamefont {Ribordy}},
  \bibinfo {author} {\bibfnamefont {W.}~\bibnamefont {Tittel}}, \ and\ \bibinfo
  {author} {\bibfnamefont {H.}~\bibnamefont {Zbinden}},\ }\href@noop {}
  {\bibfield  {journal} {\bibinfo  {journal} {Phys. Rev. A}\ }\textbf {\bibinfo
  {volume} {74}},\ \bibinfo {pages} {145} (\bibinfo {year} {2002})}\BibitemShut
  {NoStop}%
\bibitem [{\citenamefont {Yip}\ and\ \citenamefont {Finak}(1984)}]{Yip1984}%
  \BibitemOpen
  \bibfield  {author} {\bibinfo {author} {\bibfnamefont {G.~L.}\ \bibnamefont
  {Yip}}\ and\ \bibinfo {author} {\bibfnamefont {J.}~\bibnamefont {Finak}},\
  }\href@noop {} {\bibfield  {journal} {\bibinfo  {journal} {Opt. Lett.}\
  }\textbf {\bibinfo {volume} {9}},\ \bibinfo {pages} {423} (\bibinfo {year}
  {1984})}\BibitemShut {NoStop}%
\bibitem [{\citenamefont {Miliou}\ \emph {et~al.}(1991)\citenamefont {Miliou},
  \citenamefont {Srivastava},\ and\ \citenamefont {Ramaswamy}}]{Miliou:91}%
  \BibitemOpen
  \bibfield  {author} {\bibinfo {author} {\bibfnamefont {A.~N.}\ \bibnamefont
  {Miliou}}, \bibinfo {author} {\bibfnamefont {R.}~\bibnamefont {Srivastava}},
  \ and\ \bibinfo {author} {\bibfnamefont {R.~V.}\ \bibnamefont {Ramaswamy}},\
  }\href {\doibase 10.1364/AO.30.000674} {\bibfield  {journal} {\bibinfo
  {journal} {Appl. Opt.}\ }\textbf {\bibinfo {volume} {30}},\ \bibinfo {pages}
  {674} (\bibinfo {year} {1991})}\BibitemShut {NoStop}%
\bibitem [{\citenamefont {Li$\tilde{\rm n}$ares}\ \emph
  {et~al.}(1994)\citenamefont {Li$\tilde{\rm n}$ares}, \citenamefont {Montero},
  \citenamefont {Prieto},\ and\ \citenamefont {de~La~Fuente}}]{Linares1994}%
  \BibitemOpen
  \bibfield  {author} {\bibinfo {author} {\bibfnamefont {J.}~\bibnamefont
  {Li$\tilde{\rm n}$ares}}, \bibinfo {author} {\bibfnamefont {C.}~\bibnamefont
  {Montero}}, \bibinfo {author} {\bibfnamefont {X.}~\bibnamefont {Prieto}}, \
  and\ \bibinfo {author} {\bibfnamefont {R.}~\bibnamefont {de~La~Fuente}},\
  }\href@noop {} {\bibfield  {journal} {\bibinfo  {journal} {J. Mod. Opt.}\
  }\textbf {\bibinfo {volume} {41}},\ \bibinfo {pages} {5} (\bibinfo {year}
  {1994})}\BibitemShut {NoStop}%
\bibitem [{\citenamefont {Tervonen}\ \emph {et~al.}(2011)\citenamefont
  {Tervonen}, \citenamefont {West},\ and\ \citenamefont
  {Honkanen}}]{Tervonen11}%
  \BibitemOpen
  \bibfield  {author} {\bibinfo {author} {\bibfnamefont {A.}~\bibnamefont
  {Tervonen}}, \bibinfo {author} {\bibfnamefont {B.~R.}\ \bibnamefont {West}},
  \ and\ \bibinfo {author} {\bibfnamefont {S.}~\bibnamefont {Honkanen}},\
  }\href@noop {} {\bibfield  {journal} {\bibinfo  {journal} {Opt. Engin.}\
  }\textbf {\bibinfo {volume} {50}},\ \bibinfo {pages} {071107} (\bibinfo
  {year} {2011})}\BibitemShut {NoStop}%
\bibitem [{\citenamefont {Li$\tilde{\rm n}$ares}\ \emph
  {et~al.}(2000)\citenamefont {Li$\tilde{\rm n}$ares}, \citenamefont {Moreno},\
  and\ \citenamefont {Nistal}}]{Linares2000JMO}%
  \BibitemOpen
  \bibfield  {author} {\bibinfo {author} {\bibfnamefont {J.}~\bibnamefont
  {Li$\tilde{\rm n}$ares}}, \bibinfo {author} {\bibfnamefont {V.}~\bibnamefont
  {Moreno}}, \ and\ \bibinfo {author} {\bibfnamefont {M.~C.}\ \bibnamefont
  {Nistal}},\ }\href@noop {} {\bibfield  {journal} {\bibinfo  {journal} {J.
  Mod. Opt.}\ }\textbf {\bibinfo {volume} {47}},\ \bibinfo {pages} {599}
  (\bibinfo {year} {2000})}\BibitemShut {NoStop}%
\bibitem [{\citenamefont {Li$\tilde{\rm n}$ares}\ \emph
  {et~al.}(2001)\citenamefont {Li$\tilde{\rm n}$ares}, \citenamefont {Moreno},
  \citenamefont {Nistal},\ and\ \citenamefont {Salgueiro}}]{Linares2001JMO}%
  \BibitemOpen
  \bibfield  {author} {\bibinfo {author} {\bibfnamefont {J.}~\bibnamefont
  {Li$\tilde{\rm n}$ares}}, \bibinfo {author} {\bibfnamefont {V.}~\bibnamefont
  {Moreno}}, \bibinfo {author} {\bibfnamefont {M.~C.}\ \bibnamefont {Nistal}},
  \ and\ \bibinfo {author} {\bibfnamefont {J.~R.}\ \bibnamefont {Salgueiro}},\
  }\href@noop {} {\bibfield  {journal} {\bibinfo  {journal} {J. Mod. Opt.}\
  }\textbf {\bibinfo {volume} {48}},\ \bibinfo {pages} {789} (\bibinfo {year}
  {2001})}\BibitemShut {NoStop}%
\bibitem [{\citenamefont {Walker}\ and\ \citenamefont
  {Wilkinson}(1983)}]{Walker:83}%
  \BibitemOpen
  \bibfield  {author} {\bibinfo {author} {\bibfnamefont {R.~G.}\ \bibnamefont
  {Walker}}\ and\ \bibinfo {author} {\bibfnamefont {C.~D.~W.}\ \bibnamefont
  {Wilkinson}},\ }\href {\doibase 10.1364/AO.22.001929} {\bibfield  {journal}
  {\bibinfo  {journal} {Appl. Opt.}\ }\textbf {\bibinfo {volume} {22}},\
  \bibinfo {pages} {1929} (\bibinfo {year} {1983})}\BibitemShut {NoStop}%
\bibitem [{\citenamefont {Minford}\ \emph {et~al.}(1994)\citenamefont
  {Minford}, \citenamefont {Korotky},\ and\ \citenamefont
  {Alferness}}]{Minford1982}%
  \BibitemOpen
  \bibfield  {author} {\bibinfo {author} {\bibfnamefont {W.~J.}\ \bibnamefont
  {Minford}}, \bibinfo {author} {\bibfnamefont {S.~K.}\ \bibnamefont
  {Korotky}}, \ and\ \bibinfo {author} {\bibfnamefont {R.~C.}\ \bibnamefont
  {Alferness}},\ }\href@noop {} {\bibfield  {journal} {\bibinfo  {journal}
  {IEEE J. Quantum Electron.}\ }\textbf {\bibinfo {volume} {QE-18}},\ \bibinfo
  {pages} {1802} (\bibinfo {year} {1994})}\BibitemShut {NoStop}%
\bibitem [{\citenamefont {Prieto-Blanco}\ and\ \citenamefont {J.Li$\tilde{\rm
  n}$ares}(2012)}]{Prieto2012}%
  \BibitemOpen
  \bibfield  {author} {\bibinfo {author} {\bibfnamefont {X.}~\bibnamefont
  {Prieto-Blanco}}\ and\ \bibinfo {author} {\bibnamefont {J.Li$\tilde{\rm
  n}$ares}},\ }\href@noop {} {\bibfield  {journal} {\bibinfo  {journal} {IEEE
  Photonics J.}\ }\textbf {\bibinfo {volume} {4}},\ \bibinfo {pages} {65}
  (\bibinfo {year} {2012})}\BibitemShut {NoStop}%
\bibitem [{\citenamefont {Montero-Orille}\ \emph {et~al.}(2020)\citenamefont
  {Montero-Orille}, \citenamefont {Gonz\'alez-N\'u$\tilde{\rm n}$ez},
  \citenamefont {Prieto-Blanco}, \citenamefont {Moreno}, \citenamefont
  {Mouriz}, \citenamefont {Nistal},\ and\ \citenamefont {Li$\tilde{\rm
  n}$ares}}]{Montero2020}%
  \BibitemOpen
  \bibfield  {author} {\bibinfo {author} {\bibfnamefont {C.}~\bibnamefont
  {Montero-Orille}}, \bibinfo {author} {\bibfnamefont {H.}~\bibnamefont
  {Gonz\'alez-N\'u$\tilde{\rm n}$ez}}, \bibinfo {author} {\bibfnamefont
  {X.}~\bibnamefont {Prieto-Blanco}}, \bibinfo {author} {\bibfnamefont
  {V.}~\bibnamefont {Moreno}}, \bibinfo {author} {\bibfnamefont
  {D.}~\bibnamefont {Mouriz}}, \bibinfo {author} {\bibfnamefont {M.~C.}\
  \bibnamefont {Nistal}}, \ and\ \bibinfo {author} {\bibfnamefont
  {J.}~\bibnamefont {Li$\tilde{\rm n}$ares}},\ }\href@noop {} {\bibfield
  {journal} {\bibinfo  {journal} {EPJ Web Conf.}\ }\textbf {\bibinfo {volume}
  {238}},\ \bibinfo {pages} {03006} (\bibinfo {year} {2020})}\BibitemShut
  {NoStop}%
\end{thebibliography}%

\end{document}